\documentclass[epj]{svjour}
\usepackage{graphics}
\begin{document}
\title{ Beta Decays With  Momentum Space Majorana Spinors}
\author{M. Kirchbach\inst{1} \and C. Compean\inst{1}
\and L. Noriega\inst{2}
}                     
%
%
\institute{
Instituto de Fisica, UASLP,
Av. Manuel Nava 6, Zona Universitaria,\\
San Luis Potosi, SLP 78290, M\'exico
\and  Facultad de Fisica, UAZ,
Av. Preparatoria 301, Fr. Progreso,\\
Zacatecas, ZAC 98062, M\'exico
}
\date{Received: 11 November 2003 / Revised version: 20 April 2004}
%
\abstract{
We construct and apply to $\beta $ decays a
truly neutral local quantum field that is 
entirely based upon momentum space Majorana spinors. 
We make the observation that theory with momentum space
Majorana spinors of real $C$ parities is equivalent to Dirac's theory.
{}For imaginary $C$ parities, the neutrino mass 
can drop out from the single $\beta $ decay trace and 
reappear in $0\nu \beta\beta $, a curious and in principle  
experimentally testable signature for a  non-trivial impact of 
Majorana framework in experiments with polarized sources.
\PACS{
      {PACS-key}{11.30.Er Charge conjugation, parity, time reversal, 
and other discrete symmetries- 14.60.St.Non-standard-model neutrinos, 
right handed-neutrinos, etc.} 
     } 
} 
\maketitle
%

\def\beq{\begin{eqnarray}}
\def\eeq{\end{eqnarray}}


\def\s{\mbox{\boldmath$\displaystyle\mathbf{\sigma}$}}
\def\J{\mbox{\boldmath$\displaystyle\mathbf{J}$}}
\def\K{\mbox{\boldmath$\displaystyle\mathbf{K}$}}
\def\P{\mbox{\boldmath$\displaystyle\mathbf{P}$}}
\def\p{\mbox{\boldmath$\displaystyle\mathbf{p}$}}
\def\hp{\mbox{\boldmath$\displaystyle\mathbf{\widehat{\p}}$}}
\def\x{\mbox{\boldmath$\displaystyle\mathbf{x}$}}
\def\0{\mbox{\boldmath$\displaystyle\mathbf{0}$}}
\def\bv{\mbox{\boldmath$\displaystyle\mathbf{\varphi}$}}
\def\hbv{\mbox{\boldmath$\displaystyle\mathbf{\widehat\varphi}$}}

\def\bg{\mbox{\boldmath$\displaystyle\mathbf{\gamma }$}}

\def\bl{\mbox{\boldmath$\displaystyle\mathbf{\lambda}$}}
\def\br{\mbox{\boldmath$\displaystyle\mathbf{\rho}$}}
\def\1{\mbox{\boldmath$\displaystyle\mathbf{1}$}}
\def\bfhh{\mbox{\boldmath$\displaystyle\mathbf{(1/2,0)\oplus(0,1/2)}\,\,$}}

\def\mn{\mbox{\boldmath$\displaystyle\mathbf{\nu}$}}
\def\amn{\mbox{\boldmath$\displaystyle\mathbf{\overline{\nu}}$}}

\def\mne{\mbox{\boldmath$\displaystyle\mathbf{\nu_e}$}}
\def\amne{\mbox{\boldmath$\displaystyle\mathbf{\overline{\nu}_e}$}}
\def\rlh{\mbox{\boldmath$\displaystyle\mathbf{\rightleftharpoons}$}}

\def\wm{\mbox{\boldmath$\displaystyle\mathbf{W^-}$}}
\def\hh{\mbox{\boldmath$\displaystyle\mathbf{(1/2,1/2)}$}}
\def\h00h{\mbox{\boldmath$\displaystyle\mathbf{(1/2,0)\oplus(0,1/2)}$}}
\def\znbb{\mbox{\boldmath$\displaystyle\mathbf{0\nu \beta\beta}$}}



\vspace{1truecm}

\section{Introduction.}  
The theory of truly neutral fermions is based upon quantum fields
that are $C$ eigenstates.
In the convention of Ref.~\cite{Peskin} the charge conjugation operator
reads
\beq
C= i\gamma_2 K\, ,
\label{C_C_O}
\eeq 
with $K$ standing for the operation of complex conjugation.
The calculus of widest use for neutral spin 1/2 fermions
is based upon a field that is the sum,
\beq
\nu_M (x) =\frac{1}{\sqrt{2}}\left( \Psi_D (x)+\Psi_D (x)^c\right)\, ,
\label{Major_F}
\eeq 
of a Dirac quantum field, $\Psi_D (x)$, and its charged-conjugate,
$\Psi_D (x)^c$,  \cite{em1937},\cite{Kaiser},\cite{Commins}.
This so called Majorana quantum field (denoted by $\nu_M (x)$ ) is given by
\beq
\nu_M (x)=
\int \frac{d^3\p}{2 p_0(2\pi)^\frac{3}{2}} &&
\sum_{h}
{\Big[} 
u_h(\p ) a_{ h }(\p) e^{-i p\cdot x}\nonumber\\          
&+& 
v_h(\p)\,\,\,
\lbrack \lambda\, a^\dagger_{h} (\p)\rbrack \, e^{i p\cdot x}
{\Big]}\, ,\nonumber\\
a_h(\p )&=&\frac{1}{\sqrt{2}}\left( b_h(\p ) + d_h^+ (\p )\right)\, ,
\label{field_opr}
\eeq
where $h=\uparrow,\downarrow $ labels spin projection,
$ b_h(\p )$, and $d_h^+ (\p )$ are in turn
fermion annihilation and anti-fermion creation operators.
The Majorana quantum field constructed in this way is of real positive
$C$ parity,
\beq
C\nu_M(x) =\nu_M(x)\, .
\label{CPsi=Psi}
\eeq
A comment on $\lambda $ is in order,
the free phase factor in the definition of
$a^\dagger_h(\p )$ in Eq.~(\ref{field_opr}).
It is known as the {\it creation phase factor\/},
was introduced in \cite{KayserPRD}, and 
secures that the phase freedom one has in the choice of
the one-partilce states does not show up in the
observables, in particular, does not change the $C$
parity of $\nu_M(x)$. It is also useful
in the construction of a real mixing matrix.\\

\noindent
Charged particle currents, 
\beq
 \left( b^+_h(\p ) +d_h(\p )\right)\bar u_h(\p ) \gamma^\mu 
\left(b_{h^\prime} (\p ) +d_{h^\prime }^+(\p ) \right)
u_{h^\prime }(\p ) \, ,
\label{delta_L_2}
\eeq
in containing the term,
$b^+ _h(\ p)\bar u_h(\p ) \gamma^\mu  d^+_{h^\prime }(\p ) u_{h^\prime }(\p )$,
allow the lepton number to change by two units, $|\Delta L=2|$,
and account for the neutrinoless double $\beta $ decay, $0\nu \beta \beta $,
a process in which we are particularly interested here.
The resulting $0\nu\beta \beta $ trace 
is expressed in terms of momentum space Dirac spinors, $u_h(\p )$, 
and $v_h(\p )$. Recall, that momentum
space Dirac  spinors diagonalize the parity operator, 
$P{\mathcal R}$ with ${\mathcal R}$ standing for space 
reflection,
\beq
P u_h(-\p )=\eta_1^* u_h(\p )\, , \quad
P v_h(-\p )=\eta_2^* v_h(\p )\, . 
\label{par_Dir}
\eeq
The spatial parity of the Dirac spinors has been denoted 
by $\eta^* _j$ with $j=1,2$ and $\eta_j\eta_j^*=1$, and can be either
real or, pure imaginary. 
Dirac spinors with real spatial parity, 
$P=\gamma_0$, correspond to a real mass, and are of common use.
Those with a pure imaginary spatial parity, $P=\gamma_0K$, 
correspond to imaginary mass and are ruled out because of acausal 
propagation (see Ref.~\cite{MCN} for details).
To recapitulate, the Majorana quantum field is constructed as an 
afterthought of the Dirac quantum field.

\noindent
On the other hand, one can have also  momentum space  
spinors, here denoted by $\Psi_{M }^{h; (\epsilon_j)} (\p )$, 
that have the property to diagonalize the charge 
conjugation operator,
\beq
i\gamma_2 \left( \Psi_{M}^{h;(\epsilon_j)}(\p )\right) ^*=
\epsilon_j^* 
\Psi_{M}^{h;(\epsilon_j)}(\p )\,, \,\, \epsilon_j\epsilon_j^*=1\,,
\,\, j=1,2\, .
\label{Maj_p_C}
\eeq
Such spinors are referred to as momentum space Majorana spinors
\cite{Peskin}, \cite{Kaiser}, \cite{Ramon}, \cite{Ria_Fay}, \cite{Pokorski}, 
and find mentioning in neutrino oscillations 
\cite{Bilenky}, \cite{Esposito}.\\

\noindent
We now ask the question whether
$C$ parity spinors qualify for the construction of
a truly neutral local quantum field, and without reference to the
Dirac quantum field, i.e. a field that is distinct from
Eq.~(\ref{Major_F}).
It is the goal of the present study to 
design such a field and compare it to $\nu _M(x)$.\\

\noindent
To do so we follow the standard textbook quantization 
procedure, and construct as a first step 
$\Psi_M^{h;(\epsilon_j)}(\p )$ projectors and propagators. 
Here we run into the first obstacle.
Because of non-commutativity of $\gamma_0$ and $\gamma_2$, 
Eqs.~(\ref{par_Dir}), and (\ref{Maj_p_C}) can not be diagonalized
by same set of solutions. Momentum space Majorana spinors are
linear combinations of Dirac $u_h(\p )$ and $v_h(\p )$ spinors
and satisfy a {\it system of two coupled\/}  Dirac like equations.
An appropriate technique to treat propagators resulting from 
systems of two coupled spinor equations is to 
(i) first organize the two spinors in one
auxiliary eight dimensional, $(8d)$, spinor,
(ii) then construct associated projectors,
(iii) next obtain from them the propagators, and 
(iv) carry out the quantization procedure,
a program realized in Section 2 below.

\noindent
We consider two types of solutions to
Eq.~(\ref{Maj_p_C}), one with real,
the other with imaginary $C$ parities.
Naively one could expect Majorana spinors of imaginary $C$ parity 
to propagate acausally, similarly as imaginary spatial 
parity Dirac spinors.
As we shall see below, this is not the case because 
for coupled Majorana spinor equations  there is no immediate 
relation between $C$ parity and causality. 
In the auxiliary space we build spinors of 
real masses and causal propagators for any $C$ parity of the 
underlying Majorana spinors, and exploit them for the
construction of local quantum fields. We use these fields
in the calculation of $\beta $ decays.
The $(8d)$ space considered by us is in its nature auxiliary because
physics observables related to baryon $\beta $ decays
depend on traces, and our $(8d)$ traces always reduce to four 
dimensional traces expressed in terms of Dirac spinors.
At that level we can compare  Majorana  and Dirac frameworks.
We show that single $\beta $ decays of polarized sources
distinguish between Majorana and Dirac momentum
space spinors, a result discussed in Section 3 below.

\noindent
The paper is organized as follows. In the next Section we
compare Dirac and Majorana momentum space spinors and obtain
coupled equations for  Majorana spinors. 
Sections 3 and 4 are in turn devoted to
single $\beta$ and double $0\nu \beta\beta $ decays.
The main text closes with a brief Summary.

\section{Dirac versus Majorana momentum space spinors.}
The generic $C$ parity spinors can be written as
\begin{eqnarray}
\Psi^{h; (\epsilon_j)}_{M}=
\left(
\begin{array}{c}
\epsilon_{j} \xi_{1}^*\\
\epsilon_{j} \xi_{ 2}^*\\
\xi^1\\
\xi^2
\end{array}\right)\, , &\quad& \xi_\alpha^*=
(i\sigma_2)_{\alpha\beta}\left( \xi^\beta\,\right)^*\, ,
\nonumber\\
\dot \zeta = \left(
\begin{array}{c}
\xi_{1}^*\\
\xi_{ 2}^*\\
\end{array}
\right)\simeq (\frac{1}{2},0)\, ,
&\quad & 
\zeta = \left(
\begin{array}{c}
\xi^{1}\\
\xi^{ 2}\\
\end{array}
\right)\simeq (0,\frac{1}{2})\, .
\label{Maj_1}
\end{eqnarray}
Here, $\xi^\alpha  $  and $\xi_\beta^*$ are the complex components
of $(0,1/2)$, and $(1/2,0)$, respectively, which in turn correspond 
to spinor- , and co-spinor, while $i\sigma_2$, with $\sigma_2 $ 
standing for the second Pauli matrix, plays the role of metric in 
spinor space \cite{Hladek}. Note that for charged Dirac spinors, 
$(1/2,0)$, and  $(0,1/2)$ are uncorrelated. 

\noindent
As long as parity-- and charge-conjugation operators in
$(1/2,0)\oplus (0,1/2)$ do not commute,
$\Psi_M^{h ;(\pm 1)}(\p )$ will be
a linear combination of Dirac's $u_h(\p )$ and $v_h(\p )$ spinors,
and visa versa.  The easiest way to find the linear combination is to solve
Eqs.~(\ref{par_Dir}), and (\ref{Maj_p_C}) in the rest frame,
and compare the solutions. 
To be specific, we exploit Cartesian rest frame spinors,
here denoted by $\zeta_h(\0 )\simeq (0,1/2)$,
\beq
\zeta_\uparrow (\0 ) =
\sqrt{m}\left(
\begin{array}{c}
1\\
0
\end{array}
\right)\, ,
\qquad
\zeta_\downarrow (\0) =
\sqrt{m}\left(
\begin{array}{c}
0\\
-1
\end{array}
\right)\,.
\label{Cartesian}
\eeq 

\subsection{Momentum space Majorana spinors of real $C$ parity and
symmetric Majorana mass term.}
{}For concreteness, we first consider real $C$ parity spinors,
i.e. $\epsilon_j=\pm 1$ in Eq.~(\ref{Maj_1}).
Next we solve Eq.~(\ref{par_Dir}), for
$u_h(\p )$, $v_h(\p )$ and Eq.~(\ref{Maj_p_C}) for 
$\Psi_M^{h;(\pm 1)}(\p )$, respectively, in following the procedure
of Ref.~\cite{KA}.
Finally, in comparing spatial-- to $C$ parity solutions
we encounter the following decomposition of 
momentum space Majorana-- into momentum space Dirac spinors:
\beq
\left(
\begin{array}{c}
\Psi_M^{ \uparrow ;(+1)  }(\p )\\
\Psi_M^{\downarrow ; (+1) }(\p )\\
\Psi_M^{ \uparrow ;(-1) }(\p )\\
\Psi_M^{ \downarrow ;(-1)  }(\p )
\end{array}
\right)=
\frac{1}{2}\left(
\begin{array}{cccc}
1_4 &  1_4 & -1_4 &  1_4 \\
-1_4 & 1_4 & - 1_4 & -1_4 \\
1_4 & -1_4 & -1_4 & -1_4 \\
1_4 & 1_4 & 1_4 & -1_4 
\end{array}
\right)\,
\left(
\begin{array}{c}
u_{\uparrow}(\p )\\
u_{\downarrow}(\p )\\
v_{\uparrow}(\p )\\
v_{\downarrow }(\p )
\end{array}
\right)\, .\nonumber\\
\label{omega}
\eeq
Notice unitarity of the transformation matrix.\\

\noindent
{}From the last equation one immediately reads off that
Majorana spinors are self-orthogonal. Row by row one finds,
\beq
\overline{\Psi}_M^{h ;(\epsilon_j)}(\p )\Psi_M^{h ;(\epsilon_j )}(\p )
&=&
\sum_{h=\uparrow, \downarrow} \bar u_h (\p)u_h(\p )\nonumber\\
&+&\sum_{h=\uparrow, \downarrow} \bar v_h (\p)v_h(\p )
=0\, ,
\label{self_orth}
\eeq
where we used $\bar u_h(\p )u_h(\p )=2m$, and 
$\bar v_h(\p )v_h(\p )=-2m$.
Moreover, the $\Psi_M^{h;(\pm )}(\p )$'s
are cross-normalized according to
\beq
\overline{\Psi}_M^{\uparrow;(+1)}(\p )\Psi_M^{\downarrow ;(-1)}(\p )&=&
\overline{\Psi}_M^{\downarrow;(+1)}(\p )\Psi_M^{\uparrow ;(-1)}(\p ) =
2m\, ,\nonumber\\
\overline{ \Psi}_M^{\downarrow;(+1)}(\p )\Psi_M^{\uparrow;(-1)}(\p ) &=&
\overline{ \Psi}_M^{\uparrow;(-1)}(\p )\Psi_M^{\downarrow;(+1)}(\p )= -2m 
\label{cross_n}\nonumber\\
\eeq
Self-orthogonality and cross-normalization are
unpleasant properties as they frustrate covariant
propagation and local canonical quantization (see Ref.~(\cite{MCN} for
technical details). 
It is one of the goals of the present study to find a way out of 
these problems.

\noindent
The equation satisfied by the momentum space Majorana spinors
is now determined  in subjecting 
$\left[ (p\!\!\!/-m)\otimes 1_2\right] \oplus
 \left[ (p\!\!\!/+m)\otimes 1_2\right]$ to a similarity 
transformation by means of the matrix in the rhs 
in Eq.~(\ref{omega}):
\beq
\frac{1}{4}&&\left(
\begin{array}{cccc}
1_4 &  1_4 & -1_4 &  1_4 \\
-1_4 & 1_4 & - 1_4 & -1_4 \\
1_4 & -1_4 & -1_4 & -1_4 \\
1_4 & 1_4 & 1_4 & -1_4 
\end{array}
\right)
\left(
\begin{array}{cccc}
p\!\!\!/ -m & 0_4&0_4 &0_4\\
0_4&p\!\!\!/ -m &0_4&0_4\\
0_4&0_4&p\!\!\!/ +m &0_4\\
0_4&0_4&0_4&p\!\!\!/ +m 
\end{array}
\right)\nonumber\\
&&\left(\begin{array}{cccc}
1_4 &  1_4 & -1_4 &  1_4 \\
-1_4 & 1_4 & - 1_4 & -1_4 \\
1_4 & -1_4 & -1_4 & -1_4 \\
1_4 & 1_4 & 1_4 & -1_4 
\end{array}
\right)^{-1}\nonumber\\
&=&\left(
\begin{array}{cccc}
p\!\!\!/&0_4&0_4&m1_4\\
0_4&p\!\!\!/&-m1_4&0_4\\
0_4&-m1_4&p\!\!\!/&0_4\\
m1_4&0_4&0_4&p\!\!\!/
\end{array}
\right).
\label{sim_trans_1}
\eeq
The resulting set of equations for momentum space Majorana
spinors can be cast into the following block-diagonal form
\beq
\left(
\begin{array}{cccc}
p\!\!\!/ &-m1_4 & 0_4&0_4\\
-m1_4& p\!\!\!/ &0_4&0_4\\
0_4&0_4& p\!\!\!/&m1_4 \\
0_4&0_4& m1_4& p\!\!\!/ 
\end{array}
\right) 
\left(
\begin{array}{c}
\Psi_M^{ \uparrow ;(+1)  }(\p )\\
\Psi_M^{\downarrow ; (-1) }(\p )\\
\Psi_M^{ \downarrow ;(+1) }(\p )\\
\Psi_M^{ \uparrow ;(-1)  }(\p )
\end{array}
\right)=0\, .
\label{block_diag_r}
\eeq
Finally, Eq.~(\ref{block_diag_r}) is equivalently rewritten as
the following system of two coupled Dirac  equations:
\beq
\left(
\begin{array}{cc}
 p\!\!\!/ &\mp m 1_4 \\
\mp m 1_4 &p\!\!\!/ 
\end{array}
\right)
\left(
\begin{array}{c}
\Psi_M^{h ; (\epsilon_j)}(\p )\\
\Psi_M^{-h ;(-\epsilon_j)}(\p )
\end{array}
\right)
=0\, .
\label{8_space_mt}
\eeq

\noindent
At that stage it is rather instructive to recall following
properties of Dirac spinors:
\beq
\gamma_5\, u_h(\p) =v_h(\p )\, ,
&\quad &
\gamma_5\, v_h(\p) =u_h(\p )\, ,
\label{gamma_5_u_v}\\
Cu_\uparrow (\p )=v_\downarrow (\p )\, ,
&\quad& 
Cu_\downarrow (\p )=-v_\uparrow (\p )\, ,\nonumber\\
Cv_\uparrow (\p )=-u_\downarrow (\p )\, ,
&\quad& 
Cv_\downarrow (\p )=u_\uparrow (\p )\, .
\label{C_u_v}
\eeq
Insertion of Eqs.~ (\ref{gamma_5_u_v}), and (\ref{C_u_v})
into Eq.~(\ref{omega}),
allows to re-express the Majorana spinors as combinations
of the left handed (L)-- , and the charge-conjugate
right-(R) handed Dirac spinors according to
\beq
\Psi_M^{ \uparrow ;(+1)  }(\p ) &=&
u_\uparrow ^L(\p )^c +u^R_\uparrow (\p )\, ,
\nonumber\\
\Psi_M^{ \downarrow ;(+1)  }(\p )&=&
u_\downarrow ^L(\p )^c +u^R_\downarrow (\p )\, ,
\nonumber\\
\Psi_M^{ \uparrow ;(-1)  }(\p )&=&
-v_\uparrow ^R(\p ) + v^L_\uparrow (\p )^c\, ,
\nonumber\\
\Psi_M^{ \downarrow ;(-1)  }(\p )&=&
-v_\downarrow ^R(\p ) + v^L_\downarrow (\p )^c\, .
\label{link_Bilenk}
\eeq
Here,
\beq
u_h^R(\p )&=&\frac{1}{2}\left(1_4-\gamma_5\right) u_h(\p)\, ,
\nonumber\\
u_h^L(\p )^c&=& \frac{1}{2}\left(1_4 +
\gamma_5\right) i\gamma_2\, u_h^*(\p )\, ,
\label{Maj_stand}
\eeq
are same classical Majorana spinors that have been
introduced within the context of neutrino oscillations
in Refs.~\cite{Bilenky}, \cite{Esposito}. 
The two coupled Dirac-like equations (\ref{8_space_mt})
are now equivalently rewritten to
\beq
p\!\!\!/  \left( u_\uparrow ^R(\p ) +u^L_\uparrow (\p )^c\,\right)& =&
m\left( -v_\downarrow ^R(\p ) +v^L_\downarrow (\p )^c\right)\, ,\nonumber\\
p\!\!\!/  \left( 
-v_\downarrow ^R(\p ) +v^L_\downarrow (\p )^c\, \right)\, 
& =&m\left( 
u_\uparrow ^R(\p ) +u^L_\uparrow (\p )^c\,\right)\, .
\label{coupled_chann}
\eeq 
 The technique used by us to treat the
coupled equations (\ref{8_space_mt})  
is to introduce the following complete set of auxiliary 
eight dimensional spinors:
\beq
\Lambda_{l}(\p ) &= &\left(
\begin{array}{c}
u^L_\uparrow (\p )^c +u^R_\uparrow (\p )\\
\alpha_i \left(-v_\downarrow^R(\p ) +v^L_\downarrow (\p )^c \right)
\end{array}
\right)\, ,\,\, l=1,7,\,\, \alpha_1=-\alpha_7=1 \, ,
\nonumber\\
\Lambda_{ k}(\p )&=&\left(
\begin{array}{c}
-v_\downarrow^R(\p ) + v^L_\downarrow (\p )^c\\
\alpha_k \left(u_\uparrow ^L(\p )^c+u_\uparrow^R(\p)      \right)
\end{array}
\right)\, ,\,\, k=2,8, \,\, \alpha_2=-\alpha_8=1\, , \nonumber\\
\Lambda _{r}(\p )&=&\left(
\begin{array}{c}
u^L_\downarrow (\p)^c + u_\downarrow^R(\p )\\
\alpha_r\left(-v_\uparrow^R(\p ) +v^L_\uparrow (\p )^c
\right)
\end{array}
\right)\, ,\,\, r=3,5,\,\, \alpha_3=-\alpha_5=-1\, , \nonumber\\
\Lambda_{s} (\p )&=&\left(
\begin{array}{c}
-v_\uparrow^R(\p ) +v^L_\uparrow (\p )^c \\
\alpha_s \left(u^L_\downarrow (\p)^c + u_\downarrow^R(\p )  \right)
\end{array}
\right)\, ,\,\, s=4,6,\,\, \alpha_4=-\alpha_6=-1\, .
\nonumber\\
\label{d_space}
\eeq
The advantage of the auxiliary spinors is that
they can be ortho-normalized provided, one exploits 
the matrix from the mass term
in Eq.~(\ref{8_space_mt}) as a metric in the 
auxiliary space and defines $\bar \Lambda_k(\p )$ as
\beq
\bar \Lambda_{k}(\p )&=& \lbrack \Lambda_{k}(\p )\rbrack \, 
^\dagger \, \Gamma_8\,\Gamma^0 , \quad  k=1,...,8,\nonumber\\
\Gamma_0=\gamma_0\otimes 1_2\, ,
&\,\,&\Gamma_8 =\left(
\begin{array}{cc}
0_4&1_4\\
1_4&0_4
\end{array}
\right)\,. 
\label{metric_8_real}
\eeq
With this definition, the norms of the $\Lambda_j(\p )$ 
spinors are obtained as
\beq
\bar \Lambda_{i}(\p )\Lambda_{i}(\p )&=& +4m\, ,\quad
 i=1,2,7,8,
\nonumber\\
\bar \Lambda_{j}(\p )\Lambda_{j}(\p )&=&-4m\, ,
\quad j=3,4,5,6\,, \nonumber\\
\bar \Lambda_k(\p )\Lambda_l(\p) &=&0\, , \quad k\not=l\, .
\label{m8_real}
\eeq
It is interesting to express 
$\bar \Lambda_i(\p )\Lambda_i(\p )$ in terms of 
$u_h^R(\p )$, $u^L_h(\p )^c$, $v_h^R(\p )$, and $v_h^L(\p )^c$.
To be specific, for $i=1$ we find
\beq
\bar \Lambda_1(\p)\Lambda_1(\p )&=&
-\overline{ v^R_\downarrow (\p )} u_\uparrow ^L(\p )^c
-\left( \overline{ v_\downarrow^R(\p )} u_\uparrow ^L(\p )^c\right)^\dagger
\nonumber\\
&+& \overline{ v^L_\downarrow }(\p )^c u^R_\uparrow (\p )
+ \left( \overline{v^L_\downarrow }(\p )^c 
u^R_\uparrow (\p )\right)^\dagger \, .
\label{Maj_mass_term}
\eeq
In the standard notations of Refs.~\cite{Bilenky}, \cite{Esposito},
the latter equation translates into
a Majorana mass term with a real symmetric mass
matrix, $\Gamma^8$,  in the space of the states
\begin{equation}
\left(
\begin{array}{c}
\nu^c_h\, ^L + \nu^R_h\, \\
\\
\pm \left( -\bar \nu ^R_{-h} + \bar \nu ^c_{-h }\, ^L\right)
\end{array}
\right)\, ,
\label{Maj_mt_symm}
\end{equation}
describing one neutrino-generation.

\noindent
Equation (\ref{m8_real}) shows that the auxiliary $(8d)$ space
contains equal numbers of spinors of real positive--,  
and of real negative norms, much alike the Dirac space.
This advantage allows for a
canonical quantization {\it \'a la\/} Dirac 
when introducing the  {\it local \/}  $\Psi_{\lbrace 8\rbrace }(x)$ 
field operator as 
\beq
\Psi_{\lbrace 8\rbrace } (x)=
\int dV &&
{\Big[ }
\sum_{k=1,2, 7,8}
\Lambda_{k} (\p ) a_{k}(\p )\, e^{-i p\cdot x}\nonumber\\
&+&
\sum_{j=3,4,5,6 }\Lambda_{j}(\p ) a^\dagger _{j} (\p ) 
\, e^{i p\cdot x}
{\Big] }\, .
\label{Maj_field_opr}
\eeq
Here, $dV$ is the appropriate phase volume.
This local quantum field is built on top of momentum space
Majorana spinors, and the counterpart of Eq.~(\ref{field_opr}).
It allows to calculate $\beta $ decays in terms of 
$\Lambda_i(\p )$ momentum space spinors. 

\subsection{Momentum space Majorana spinors of pure imaginary $C$ parity
and anti-symmetric Majorana mass term.} 
{}For momentum space Majorana spinors of pure imaginary
$C$ parity, $\epsilon_j^*=\mp i$, the transformation
matrix in Eq.~(\ref{omega}) changes to   
\beq
\left(
\begin{array}{cccc}
1_4 &  1_4 & -1_4 &  1_4 \\
-1_4 & 1_4 & - 1_4 & -1_4 \\
1_4 & -1_4 & -1_4 & -1_4 \\
1_4 & 1_4 & 1_4 & -1_4 
\end{array}
\right)\,\longrightarrow
\left(
\begin{array}{cccc}
1_4 &  -i1_4 & -1_4 & -i 1_4 \\
i1_4 & 1_4 & i 1_4 & -1_4 \\
1_4 & i1_4 & -1_4 & i1_4 \\
-i1_4 & 1_4 & -i 1_4 & -1_4 
\end{array}
\right)\, .\nonumber\\
\label{omega_im}
\eeq
As a result, in place of Eq.~(\ref{8_space_mt}), 
one finds
\beq
\left(
\begin{array}{cc}
p\!\!\!/&\mp im 1_4   \\
\pm i m1_4&  p\!\!\!/  
\end{array}
\right)
\left(
\begin{array}{c}
\Psi_M^{ \uparrow ; (\mp i) }(\p) \\
\Psi_M^{ \downarrow ;(\mp i) }(\p) \\
\end{array}
\right)&=&0\, .
\label{Maj_Dirac}
\eeq
In nullifying the determinant of the latter equation,
one obtains the standard time-like energy momentum dispersion
relation, $p^2-m^2=0$, and delivers thereby the proof that 
imaginary $C$ parity, contrary to imaginary spatial parity,
does not necessarily imply acausal  spinor propagation. 
Also these spinors are self-orthogonal
\beq
\overline{\Psi}_M^{h; (\mp i)}(\p) \Psi_M^{h;(\mp i)} (\p) &=& 0\, ,
\label{self_orth_im}
\eeq
and cross-normalized according to
\beq
\overline{\Psi}_M^{h; (\mp i)}(\p) \Psi_M^{-h;(\mp i)} (\p) 
&=& \pm  2 i m
(\delta _{h\uparrow}-\delta _{h\downarrow})\, ,
\label{bi_orth_im}
\eeq
a property termed to as  {\it bi-orthogonality\/} in Refs.~\cite{Ah2}.
Notice that the imaginary cross-norms change sign upon reversing the
order of the spinors.
At the present stage this may look odd
but in the long term it will be of interest in so far as
it will amount to  slightly different physics
compared to the real  $C$ parity Majorana spinors in Eq.~(\ref{Maj_1}).
The coupled equations (\ref{Maj_Dirac}) 
have been written down (up to notational differences)
already in Ref.~\cite{VVD1995} by inspection of  
explicitly constructed momentum space Majorana spinors.

\noindent
The complete set of auxiliary $(8d)$ spinors corresponding to 
Eqs.~(\ref{Maj_Dirac}) is introduced as
\beq
\Lambda^\tau  _1(\p )&=&\left(
\begin{array}{c}
u_\uparrow ^R(\p) \mp iu^L_\uparrow (\p )^c\\
\eta_1
\left(u_\downarrow ^R(\p) \mp iu^L_\downarrow (\p )^c\right)
\end{array}
\right), \nonumber\\
\Lambda^\tau _2  (\p )&=&\left(
\begin{array}{c}
 u_\downarrow ^R(\p) \mp iu^L_\downarrow (\p )^c\\
\eta_1 \left( u_\uparrow ^R(\p) \mp iu^L_\uparrow (\p )^c\right)
\end{array}
\right), \nonumber\\
\Lambda ^\tau  _3(\p )&=&\left(
\begin{array}{c}
-v_\uparrow^R(\p) \pm i v_\uparrow^L(\p )^c\\
\eta_2 \left(-v_\downarrow^R(\p) \pm i v_\downarrow^L(\p )^c     \right)
\end{array}
\right), \nonumber\\
\Lambda^\tau _ 4 (\p )&=&\left(
\begin{array}{c}
  -v_\downarrow^R(\p) \pm i v_\downarrow^L(\p )^c          \\
\eta_2\left(-v_\uparrow^R(\p) \pm i v_\uparrow^L(\p )^c \right)
\end{array}
\right),\nonumber\\
\tau =\pm \,, &\quad& \eta_1=-\eta_2=1 \, .
\label{d_space_2}
\eeq
Defining now $\bar \Lambda^\tau_ k (\p )$ as
\beq
\bar \Lambda^\tau _k(\p )&=& 
\lbrack \Lambda^\tau _k(\p )\rbrack \, ^\dagger \,
\widetilde{\Gamma}_8\,\Gamma^0 ,\quad
\widetilde{\Gamma}_8=
\left(
\begin{array}{cc}
0_4&-i 1_4\\
i1_4 &0_4
\end{array}
\right)\, ,
\label{metric_8_imag}
\eeq
allows for the construction of an orthogonal basis in
the recent $(8d)$ space as
\beq
\bar \Lambda^\tau _j(\p )\Lambda^\tau _ j(\p )&=& + 4m\, , 
\quad \tau=+, j=1, 4\,; \quad \tau=-, j=2,3 \nonumber\\
\bar \Lambda^\tau _k (\p )\Lambda^\tau_k (\p )&=&-4m\,, 
\quad \tau=+, k=2,3\,;\quad \tau=-, k=1,4\, ,\nonumber\\
\bar \Lambda^\tau_k(\p )\Lambda_l^{\tau^\prime }(\p)&=&0\, ,
\qquad\quad\!\! \tau\not=\tau^\prime\, ,\quad k\not=l\, .
\label{m8_2}
\eeq
In terms of the degrees of freedom in Eq.~(\ref{Maj_stand}), 
say, $\bar \Lambda^-_1(\p )\Lambda^-_1(\p )$, expresses as
\beq
\bar \Lambda^-_1(\p )\Lambda_1^-(\p )&=&
- \overline{ u^L_\uparrow (\p) ^c}  u_\downarrow^R(\p )
   +\left(\overline{ u^L_\downarrow (\p) ^c}  
u_\uparrow^R(\p ) \right)^\dagger
\nonumber\\
 &-&\overline{ u^R_\downarrow} (\p ) u^L_\uparrow (\p)^c
    +(\overline{ u^R_\uparrow} (\p ) u^L_\downarrow (\p )^c)^\dagger \, .
\label{M_MT_im}
\eeq
Again, in the standard notations of Refs.~\cite{Bilenky}, \cite{Esposito},
the latter equation translates into
a Majorana mass term with an imaginary and anti-symmetric mass
matrix, $\widetilde{\Gamma}^8$, in the new space of states 
\begin{equation}
\left(
\begin{array}{c}
\nu^R_h\mp i \nu^c_h\, ^L\\
\\
 \pm \left( \nu ^R_{-h}\mp i \nu ^c_{-h }\, ^L\right)
\end{array}
\right)\, ,
\label{Maj_mt_asymm}
\end{equation}
describing one neutrino generation. 
Also this space bifurcates into equal numbers 
of spinors with real positive, and real negative norms,
much alike the Dirac space.
The matrix $\widetilde{\Gamma}_8\Gamma_0 $ plays once again 
the role of the new metric here, which this time  
is purely imaginary and anti-symmetric, which are properties that relate 
to Eq.~(\ref{bi_orth_im}).
Also here canonical quantization {\it \' a la } Dirac is straightforward.

\noindent
Comparison between Eqs.~(\ref{m8_2}) and (\ref{m8_real}) shows that
the mass matrix in the coupled equations depends on the
$C$ parity, $\epsilon_j^*$, in Eq.~(\ref{Maj_1}).
In case $\epsilon_j^*$  is real, the mass matrix  
is real and symmetric,
while in  case $\epsilon_j^*$ is pure imaginary, it
is imaginary and anti-symmetric.
Above difference reflects the difference in
the cross-normalization properties in Eqs.~(\ref{cross_n}), and 
(\ref{bi_orth_im}), respectively, and will be of pivotal importance 
in the calculation of the single beta decay performed below.

\section{ Single $\beta $ decay with momentum space Majorana spinors.}
In order to illustrate predictive power of models based upon
momentum space Majorana spinors, we  
take here a close look at single $\beta $ decay.
When one considers physical processes that
involve both  Dirac and  Majorana spinors, one
needs to  match  single- with coupled-spinor equations. 
The simplest  way to harmonize dimensions
is to amplify the Dirac spinors in analogy with 
Eqs.~(\ref{d_space}). 
In order to respect orthogonality
of $P$ eigenspinors, one has to keep spin projections
same at top and bottom. The complete set of
Dirac eight-spinors introduced in this way is given by
\beq
U_{(j;h) } (\p )=\left( 
\begin{array}{c}
u_{h }(\p )\\
\eta_j u_{h } (\p )
\end{array}
\right)\, , &\quad&
V_{(j;h )} (\p )=
\left( 
\begin{array}{c}
v_{h } (\p )\\
\eta_j v_{h } (\p )
\end{array}
\right)\, ,\nonumber\\
\eta_1 &=&-\eta_2=1\, ,
\label{8_Dir}
\eeq
respectively.  
The metric in the auxiliary Dirac space is $\Gamma_0=\gamma_0\otimes 1_2$.
To simplify notations from now on we
will suppress  the momentum, $\p $, as argument
of spinors and operators.
First we consider the auxiliary $(8d)$ space built on top
of Majorana spinors of imaginary  $C$ parity.
In order to calculate cross sections, i.e.
$(8d)$ current-current tensors, $G^{\mu\nu}$,
one has next to introduce the
eight-currents. Here we consider the interface Dirac--Majorana 
current as the $(8d)$ extension of the  Dirac  vector
current according to  
\beq
J^\mu_{(\tau ; k)\, (j;h )} &=&
\bar \Lambda^\tau _{k} \Gamma^\mu U_{(j;h )}\, ,
\nonumber\\
\Gamma^\mu &=&\gamma^\mu \otimes 1_2\,,
\quad k=1,2,7,8\, .
\label{8_currents}
\eeq
As an illustrative example, below  we rewrite, 
$J^\mu _{(+;1) (1;\uparrow) }$, in terms of the degrees 
of freedom in Eq.~(\ref{Maj_stand}) as
\begin{eqnarray*}
J^\mu _{(+;1) (1;\uparrow) }&=&
\bar\Lambda^+_1\Gamma^\mu U_{(1;\uparrow )}\nonumber\\
&=& \sum_h \overline{u_h^L}\gamma^\mu \left(u^R_{-h}\right) ^c +
\mbox{L$\leftrightarrow$R}\, .
\end{eqnarray*}

{}Mass and four-momentum of the Dirac particle will be in 
turn denoted as $m_1$, and $p_1$.
The above currents are conserved in the $m\to m_1$ limit
and have the property to 
take states $U_{(j;h )}$, of positive norm,
to  $C$ eigenstates, of positive norm too.
The current-current tensor for, say,
$J^\mu_{(\tau ; k)\, ,(j;h )}$, is calculated to be
\beq
G^{\mu \nu} &=&{1\over 2}\,
\sum_{(\tau ; k),(j;h )}{1\over 4}  
\bar \Lambda^\tau _k  \Gamma^\mu
U_{(j;h )}\,\left( \bar \Lambda^\tau _k  \Gamma^\nu
U_{(j;h )} \right)^\dagger\, .\nonumber\\
\label{8_trace_1}
\eeq
In exploiting definition of $\bar \Lambda^\tau _k $ 
in Eq.~(\ref{m8_2}) and  making use of, 
$\Gamma^\nu \, ^\dagger \Gamma^0 \, ^\dagger =\Gamma^0\Gamma^\nu $,
one finds 
\beq
G^{\mu \nu}&=&
{1\over 2} \sum_{(\tau ; k)}
{1\over 4}  
\bar \Lambda^\tau _k  \Gamma^\mu
4m_1 \Pi^D\Gamma^\nu \widetilde{\Gamma}_8^\dagger \Lambda^\tau_ k\,,
\nonumber\\
\label{8_trace_3}
4m_1 \Pi^D &=&\left(
 U_{(1;\uparrow)}\overline{U}_{(1;\uparrow)}+
U_{(2;\downarrow)}\overline{U}_{(2;\downarrow)}\right)\nonumber\\
&=&
(m_11_4 +p\!\!\!/_1)\left(
\begin{array}{cc}
 1& 1\\
1& 1
\end{array}
\right)\, .
\label{proj_Dirac_U}
\eeq
Converting Eq.~(\ref{8_trace_3}) to trace is now standard
and reads
\beq
G^{\mu \nu}
&=&{1\over 4}tr
\left(
\begin{array}{cc}
p\!\!\!/\gamma^\mu & -im \gamma^\mu \\
im \gamma^\mu &p\!\!\!/\gamma^\mu
\end{array}
\right)
\left(
\begin{array}{cc}
(p\!\!\!/_1 +m_1)\gamma^\nu  &(m_1+p\!\!\!/_1)\gamma^\nu \\
(m_1+p\!\!\!/_1)\gamma^\nu &(p\!\!\!/_1 +m_1)\gamma^\nu 
\end{array}
\right)\, 
\nonumber\\
&=&{1\over 2} tr p\!\!\!/\gamma^\mu(m_1+p\!\!\!/_1)\gamma^\nu\, .
\label{8_trace_final}
\eeq
Therefore, the trace entering the single $\beta $ decay width
turns out to be insensitive to the neutral 
fermion mass, $m$, in Eq.~(\ref{Maj_Dirac}). 

\noindent
The reason for this unexpected phenomenon is traced back
to the antisymmetric character of the cross-normalizations
in Eq.~(\ref{bi_orth_im}), and the coupled equations
(\ref{Maj_Dirac}). Above properties
show up in the trace in the form of the
anti-symmetric off diagonal matrix
$\widetilde{\Gamma}_8$ which triggers cancellation
of the neutral particle mass. 

\noindent
The drop out of the neutral lepton mass from 
the beta decay trace in Eq.~(\ref{8_trace_final}) is an interesting 
though not as dramatic a phenomenon  as the lepton masses affect
only decay traces with polarized $\beta $ decay sources (nucleon, nuclei).
Recall that the lepton masses do not show up at all in the time like
$G^{00}$,
\beq
G^{00}=2\left( E_\nu E_e +\p_\nu\cdot \p_e  +E_\nu \p_e\cdot \s
+E_e\p_\nu \cdot \s \right)\, ,
\eeq
while in the space-like $G^{ii}$ (with $i=1,2,3$) they
enter only via spin-momentum correlation terms \cite{EG}.

\noindent
Had we used momentum space Majorana spinors with a real
$C$ parity, cross-normalization and coupled equations  would be  
symmetric in accord with Eqs.~(\ref{cross_n}), and 
(\ref{8_space_mt}), respectively. In this case 
the Majorana $\beta $ decay trace would have come out identical to the
Dirac trace.
In summary, compared to Dirac phenomenology,
only momentum space Majorana spinors of imaginary $C$ parity 
allow for differences with respect to single $\beta $ 
decays of polarized sources.

\section{The neutrinoless double beta decay $\0\nu\beta\beta$.}
The neutrinoless double beta decay ($0\nu \beta \beta $) 
is a process where two neutrons in a nucleus, $A(Z,N)$, are 
converted into two protons  by the emission of two electrons
while the two antineutrinos close to a virtual internal line
(see Fig.~1)
\beq
A (Z,N) \to  A (Z+2,N-2)+ e^- + e^-\, ,
\label{0bb}
\eeq
(see Ref.~\cite{Kaiser} for details).

\begin{figure}
\resizebox{0.20\textwidth}{4.05cm}
{
  \includegraphics{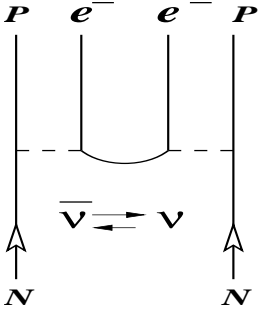}
}
\caption{ Neutrinoless double beta decay-schematic representation.}

\end{figure}

This process is associated with a second order
element of the $S$ matrix and the related amplitude,
here denoted by, $T_{0\nu\beta\beta }$, is
given by
\beq
T_{0\nu\beta\beta }=
W^\mu \, W^\eta
\lbrack \bar u_e \gamma_\mu (1+\gamma_5)u_{\nu_e}\rbrack
\lbrack \bar u_e \gamma_\eta (1+\gamma_5)u_{\nu_e}\rbrack\, .
\label{01_bb}
\eeq
In order to bring in the virtual neutrino line in Eq.~(\ref{01_bb}),
one makes use of the following identity:
\beq
\bar u_e \gamma_\eta (1+\gamma_5)u_{\nu_e}&=&
\overline{ \left( (u_e)^c\right)^c}\gamma_\eta (1+\gamma_5)
\left((u_{\nu_e})^c\right)^c\nonumber\\
&=&\bar u_{\nu_e}\, [-\gamma_\mu (1-\gamma_5)]\, v_e\, .
\label{02_bb}
\eeq
The latter expression is obtained by making use
of the relations,
$\gamma_0\gamma_\mu^\ast=\gamma_\mu\gamma_0$,
$\gamma_2\gamma_\mu=-\gamma_\mu^\ast\gamma_2$,
$\gamma_\mu^t=-\gamma_\mu$, the anti-commutation relations
between the Dirac matrices, with $''t''$ labeling the transposed.
With that Eq.~(\ref{01_bb}) takes the form
\beq
T_{0\nu\beta\beta }&=&
W^\mu \, W^\eta
\frac{1}{p_{\nu_e}^2-m_{\nu_e}^2} 
L_{\mu\eta}\, ,\nonumber\\
L_{\mu\eta}&=&
\, \bar u_e \gamma_\mu (1+\gamma_5)
\Pi^{\nu_e} \, [-\gamma_\mu (1-\gamma_5)]\, v_e \, ,\nonumber\\
\Pi^{\nu_e}&=&\sum u_{\nu_e}\bar u_{\nu_e}\, .
\label{03_bb}
\eeq
Here we suppressed $''h''$ labeling of the Dirac spinors
in order not to overload notations so that 
$\sum $ in $\Pi_{\nu_e}$ means summation over spin projections.
{}Finally, $|L_{\mu\eta}|^2$ can be converted to a
trace in the standard way as
\beq
|L_{\mu\eta }|^2 =
{\Big[}\bar u_e \gamma_\mu (1+\gamma_5)\Pi^{\nu_e}
\, \gamma_\eta (1-\gamma_5)\, v_e {\Big]}\,&& \nonumber\\
{\Big[}
\bar u_e \gamma_\lambda (1+\gamma_5)
\Pi^{\nu_e}
\, \gamma_\delta (1-\gamma_5)\, v_e {\Big]}^\dagger\,&& 
\nonumber\\
=tr{\Big[} \Pi^{u_e}\gamma_\mu (1+\gamma_5)\Pi^{\nu_e}\gamma_\eta(1-\gamma_5)
\Pi^{v_e}&&\nonumber\\
\gamma_\delta 
\gamma_0 (1+\gamma_5)\Pi^{\nu_e}\gamma_0(1-\gamma_5)\gamma_\lambda  {\Big]}
\nonumber\\
= tr{\Big[}
\Pi^{u_e}\gamma_\mu \frac{2m_{\nu_e}}{p_{\nu_e}^2}
\gamma_\eta 
(1-\gamma_5)
\Pi^{v_e} \gamma_\delta
\gamma_0 
\frac{2m_{\nu_e}}{p_{\nu_e}^2}
\gamma_0(1-\gamma_5)\gamma_\lambda{\Big]}&&
\, .\nonumber\\
\label{04_bb}
\eeq
In the latter equation the squared 
neutrino mass ($m_{\nu_e}^2$) was neglected  compared to
the squared neutrino momentum, $p_{\nu_e}^2$, with the 
well known result
\beq
(1+\gamma_5){\Pi^{\nu_e}}\gamma_\eta (1-\gamma_5)=
\frac{2m_{\nu_e}}{p_{\nu_e}^2} \gamma_\eta (1-\gamma_5)\, .
\label{Zusatz_1}
\eeq

\noindent
Now we calculate above trace within the scenario
of the previous section. To do so, one has to
perform in Eq.~(\ref{04_bb}) the replacements $\gamma_\mu\to \Gamma_\mu$,
$u_e\to U_e$, $v_e\to V_e$, $u_{\nu_e}\to \Lambda^{S/A}_ k$, and
\beq
\Pi^{\nu_e}\to
\frac{1}{2m}\left(
\begin{array}{cc}
m1_4&-ip\!\!\!/\\
ip\!\!\!/&m1_4
\end{array}
\right)
\left(
\begin{array}{cc}
0_4&-i 1_4\\
i1_4&0_4
\end{array}
\right)\, .
\label{05_bb}
\eeq
Our calculation shows that the $0\nu \beta \beta $ trace contains 
$\widetilde{\Gamma_8}^2 $ which is the 
$(8d)$ identity matrix.
In effect, one recovers  Eq.~(\ref{04_bb}) and 
the well known proportionality of the $0\nu\beta\beta $ trace 
to the square of the neutrino mass.
Therefore, the Majorana calculus does not alter results of the
Dirac theory of the neutrinoless double beta decay.

\section{Summary.}
We constructed two types of truly neutral spin-1/2
quantum fields that differ by the $C$ parity of the underlying
momentum space Majorana spinors, real versus imaginary,
a property that shows up as a difference in the
symmetry of the corresponding Majorana
mass matrices--real symmetric {\it  versus\/} imaginary 
anti-symmetric.
We exploited above fields to calculate traces of single
and neutrinoless double beta decays.
Compared to standard phenomenology,
the neutrinoless double beta decay remains unaltered
for both fields.
The result extends also to one-gaugino exchange as long as
the virtual fermion line in Fig.~1 can be  
also a massive gaugino.

\noindent
In single beta decay,
we observed a cancellation of the neutral fermion mass in the trace,
in the case of the Majorana field with the  anti-symmetric 
mass matrix.

The latter option opens the curious possibility to have a 
neutral fermion theory at hand that allows 
(polarized) tritium $\beta $ decay \cite{Bonn}  to drive the neutrino mass 
closer to zero compared to neutrino oscillation-- , and
 $0\nu\beta \beta$ measurements,
thus providing an intriguing and
in principle experimentally testable signature for 
a non-trivial  impact of momentum space Majorana spinors 
on phenomenology.

\section{Acknowledgments.}
Work supported by Consejo Nacional de Ciencia y
Tecnologia (CONACyT) Mexico under grant number C01-39820.

\end{document}